\documentclass[preprint,aps,floats,nofootinbib]{revtex4}

\usepackage{graphicx}
\usepackage{dcolumn}
\usepackage{bm}
\def\lsim{\mathrel {\vcenter {\baselineskip 0pt \kern 0pt
    \hbox{$<$} \kern 0pt \hbox{$\sim$} }}}
\def\gsim{\mathrel {\vcenter {\baselineskip 0pt \kern 0pt
    \hbox{$>$} \kern 0pt \hbox{$\sim$} }}}

\begin{document}

\title{$D-\bar{D}$ mixing constraints on FCNC with a non-universal $Z^\prime$}

\author{Xiao-Gang He}
\email{hexg@phys.ntu.edu.tw} \affiliation{ Department of Physics
and Center for Theoretical Sciences, National Taiwan University,
Taipei, Taiwan}

\author{G. Valencia}
\email[]{valencia@iastate.edu} \affiliation{Department of Physics,
Iowa State University, Ames, IA 50011}

\date{\today}

\begin{abstract}

The BaBar and Belle collaborations have recently reported evidence
for $D^0-\bar D^0$ mixing. This measurement provides the first significant constraint
on FCNC in the up-quark sector for non-universal
$Z^\prime$ models.  Attributing the observed
$D-\bar D$ mixing to new physics, we comment on the resulting rare
$D$ and $t$ decays. We also show that a CP violating semileptonic asymmetry as large
as $\sim 30\%$ is allowed by the experimental results.

\end{abstract}

\pacs{PACS numbers: 12.15.Ji, 12.15.Mm, 12.60.Cn, 13.20.Eb,
13.20.He, 14.70.Pw}

\maketitle

Mixing between a neutral meson with specific flavor and its
anti-meson is a very interesting phenomena in particle physics.
There are several systems where such mixing has been
observed \cite{pdg}: $K^0-\bar K^0$; $B^0_d -\bar B^0_d$; and
$B_s^0-\bar B^0_s$. All these cases involve mesons made of valence
down-type quarks. In principle, a neutral meson made of up-type
quarks should also exhibit mixing, as in the $D^0-\bar D^0$
system \cite{reviews}.  Evidence for $D^0$ mixing has been recently
reported, assuming CP conservation, by the BaBar \cite{babar}
\begin{eqnarray}
&&x'^2 = (-0.22\pm 0.30(stat.)\pm 0.21(syst.))\times 10^{-3}, \nonumber \\
&&y' = (9.7\pm(stat.)4.4 \pm3.1(syst.))\times 10^{-3},
\label{babarres}
\end{eqnarray}
and Belle collaborations \cite{belle}
\begin{eqnarray}
&&x=(0.80\pm 0.29(stat.)\pm 0.17(syst.))\%,\;\;y = (3.3\pm 2.4(stat.)\pm0.15(syst.))\% \nonumber \\
&&y_{CP} = (1.31\pm 0.32(stat.)\pm 0.25(syst.))\%.
\label{belleres}
\end{eqnarray}
No evidence for CP violation was observed.

Mixing in the $K^0-\bar K^0$ and $B^0_d -\bar B^0_d$ systems has
been known for quite some time. It has provided valuable
information about the Standard Model (SM) as well as very
stringent constraints on possible new physics. The first
observation of $B_s-\bar{B}_s$ mixing was reported last year by
the CDF collaboration \cite{Abulencia:2006ze}, in agreement with
the two sided bound obtained by the D0 collaboration
\cite{Abazov:2006dm}. The combined result is in agreement with the
standard model prediction at the $1\sigma$ level
\cite{Charles:2004jd}, with the measurement being slightly below
the central value of the overall standard model fit. This
measurement has been used to constrain many new physics
possibilities, including non-universal $Z^\prime$ models
\cite{He:2006bk,z-prime}.

$D^0-\bar D^0$ mixing provides an additional handle on flavor
physics within and beyond the SM. Within the SM, short distance
contributions to $D$ mixing are small \cite{reviews,Golowich:2005pt,Golowich:2006gq}.
Long distance contributions can be much larger, but they suffer from considerable
uncertainty \cite{reviews}.  The BaBar and Belle results, Eq.~\ref{babarres} and
Eq.~\ref{belleres}, can therefore be attributed to Standard Model physics,
but there is ample room for new physics contributions \cite{reviews}.

There are several scenarios beyond the SM in which it is natural
to obtain large $D^0 -\bar D^0$ mixing, such as fourth generation
models~\cite{Babu:1987xe,Hou:2006mx}, SUSY extensions of the SM
and others~\cite{reviews}. In this note we study the implications
of the $D$ mixing data on a class of non-universal $Z^\prime$
models. In these models, the flavor changing neutral currents
(FCNC) generated by tree-level $Z^\prime$ exchange induce mixing
in mesons made of down and up type quarks, albeit with arbitrary
strengths. The $D$ mixing measurement provides the first significant
constraint on the strength of these FCNC. With this input, the models allow enhancements in the
rates for rare $D$ decays which remain too small for observation,
and relatively larger rare top decays.

The parameters describing $D^0-\bar D^0$ mixing are usually
indicated by $x = \Delta m/\Gamma$ and $y=\Delta \Gamma/2 \Gamma$.
There are several ways these parameters can be determined. BaBar
measured the mixing using the wrong-sign $D^0 \to K^+ \pi^-$
decay. In this case the two directly measured parameters are $x' =
x \cos\delta_{K\pi} + y \sin\delta_{K\pi}$ and $y' =
-x\sin\delta_{K\pi} + y \cos\delta_{K\pi}$ if there is no CP
violation. Here $\delta_{K\pi}$ is the CP conserving strong phase
between the amplitudes for the doubly-Cabbibo-suppressed and the
Cabbibo-favored decays \cite{reviews}.  The Belle collaboration
measured $x$ by analyzing the Dalitz plot for $D^0 \to K^0_S
\pi^+\pi^-$, and measured $y_{CP}$ defined as, $y_{CP} =
(\tau(K^-\pi^+)/\Gamma(K^+K^-) - 1) = y\cos2\phi_D - A_m x \sin
2\phi_D$ with $A_m = 1-|q/p|$. If $\phi_D =0$, $y_{CP} = y$.

Ref.\cite{Ciuchini:2007cw} combined the BaBar \cite{babar} and Belle \cite{belle}
results to obtain the 68\% C.L. ranges for: $x= (5.5\pm 2.2)\times 10^{-3}$,
$y=(5.4\pm 2.0)\times 10^{-3}$, $\phi_D =
(0\pm 11)^\circ$, $\delta_{K\pi} = (-38\pm 46)^\circ$ and $A_m =
-0.02\pm 0.15$ which we will use in this note.

We now study the implications of the above information on
non-universal $Z^\prime$ models. The most important new effect is due to the tree-level
flavor changing interactions in the heavy $Z'$ coupling to fermions. The main effect
on $D$ mixing thus occurs through the parameter $M_{12}$ and, therefore, it predominantly
affects $x$. In this class of models there is no significant new contribution to the
parameter $\Gamma_{12}$, which is responsible for the lifetime difference $y$ \cite{Golowich:2006gq}.
We will thus assume that $\Gamma_{12}$ can be attributed to long distance standard model physics
and treat it as a parameter to be determined from the data.  Similarly, the class of models we consider,
make negligible contributions to CP violating phases in $D^0$
decay amplitudes.  We will therefore  assume that
$\Gamma_{12}$ is real. The new physics contribution to $M_{12}$, on the other hand, has a
CP violating phase in general, so we
write $M_{12} = Me^{i\phi}$. Using the well known result,
\begin{eqnarray}
\Delta m - i \Delta \Gamma/2 = 2\sqrt{(M_{12} -
i\Gamma_{12}/2)(M_{12}^*-i\Gamma^*_{12}/2)},
\end{eqnarray}
we obtain two equations
\begin{eqnarray}
4(M^2-\Gamma_{12}^2/4) = \Delta m^2 - \Delta \Gamma^2/4,\;\;4
M\Gamma_{12} \cos\phi = \Delta m \Delta \Gamma;
\end{eqnarray}
with three unknowns: $M$, $\Gamma_{12}$ and $\phi$. Therefore, the  measurement of $x$ and $y$
can not determine all the unknowns. To proceed we introduce the quantities $w = 2M/\Gamma_D$
and $z=\Gamma_{12}/\Gamma_D$, which are functions of $x$, $y$ and $\phi$. We have
\begin{eqnarray}
z^2 &=& {1\over 2} (y^2-x^2 + \sqrt{(x^2-y^2)^2 + 4 {x^2y^2\over
\cos^2\phi}}),\nonumber\\
w^2 &=& x^2-y^2 +z^2.
\end{eqnarray}

\begin{figure}[htb]
\begin{center}
\includegraphics[width=8cm]{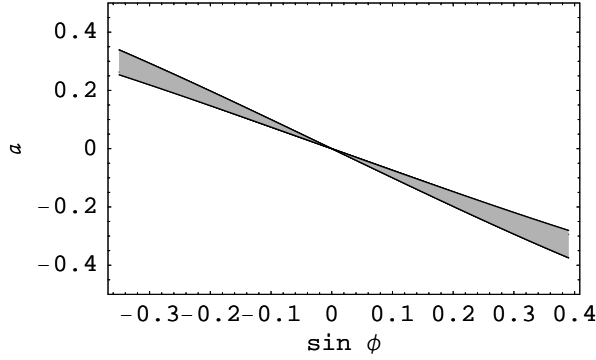}
\end{center}
\caption{Semileptonic CP violating asymmetry $a$ as a function of $\sin \phi$ for $x = (5.5 \pm 2.2) \times 10^{-3}$ and $y=(5.4\pm 2.0)\times 10^{-3}$ .} \label{fig1}
\end{figure}

Although $z$ is not known, it can be determined with a measurement of the CP violating phase $\phi$.
Already the parameter $A_m$ constrains the value of $\phi$ through the relation
\begin{equation}
A_m= {|q/p|^2-1\over |q/p|^2+1}, \;\;\;\;\left | {q\over p}\right
|^2 = \left(
\frac{z^2+w^2+2zw\sin\phi}{z^2+w^2-2zw\sin\phi}\right)^{\frac{1}{2}}.
\end{equation}
Using the central values  $x = 5.5  \times 10^{-3}$ and
$y=5.4\times 10^{-3}$ and requiring $-0.17 \leq A_m \leq 0.13$ one
finds $-0.29 \lsim \sin\phi \lsim 0.27$. If instead one allows $x$
and $y$ to vary independently in their one-sigma range one finds
$-0.39 \lsim \sin\phi \lsim 0.35$.\footnote{At three-sigma the
full range $-1 \leq \sin\phi \leq 1$ is still allowed.} For this
range, the values of $w$ and $z$ are close to those of $x$ and $y$
respectively. This phase then shows up in CP violating observables
such as the semileptonic asymmetry:
\begin{eqnarray}
a &=& {N^{-}(D^0 \to l^-X) - N^{+}(\bar D^0 \to l^+ \bar X)\over
N^{-}(D^0 \to l^-X) +
N^{+}(\bar D^0 \to l^+\bar X)}\nonumber\\
&=& {|p/q|^2-|q/p|^2\over |p/q|^2+|q/p|^2} = - {2wz\sin\phi\over
w^2+z^2}.
\end{eqnarray}
In Figure~\ref{fig1} we illustrate this asymmetry as a function of $\sin\phi$  for $x = (5.5 \pm 2.2) \times 10^{-3}$, $y=(5.4\pm 2.0)\times 10^{-3}$ and the range for $\sin\phi$ allowed by $A_m$. We emphasize
that the above analysis applies to any model where CP violation  appears only in the mixing parameter $M_{12}$.

Next we review the relevant features for meson mixing specific to non-universal $Z'$ models starting from a generic
parametrization for the $Z'$ coupling to quarks,
\begin{eqnarray}
L = {g\over 2 c_W} \bar q_i \gamma^\mu \left ( a_{ij} P_L + b_{ij}
P_R\right ) q_j Z'_\mu.
\end{eqnarray}
The exchange of $Z'$ at tree-level, generates a $\Delta f=2$ effective
Hamiltonian that can induce mixing. This was
discussed in detail in Ref.~\cite{He:2006bk} for $B-\bar B$
mixing. To apply that result to the case of $D$-mixing we only
need to continue the QCD running down to the charm scale and
concentrate on the terms with $\Delta c =2$, i.e. $ij=cu$,
\begin{eqnarray}
M^{P,Z'}_{12} &=& {G_F\over \sqrt{2}}{m^2_Z\over
m^2_{Z^\prime}}{1\over 3} f_P^2 M_P B_P
\eta_{Z'}^{6/21}\eta_{b}^{6/23}\eta_D^{6/25}\left ( a^2_{ij}+b^2_{ij}
+\eta_{Z'}^{-3/21}\eta_{b}^{-3/23}\eta_D^{-3/25}{1\over 2} a_{ij}b_{ij}(2\epsilon -3)
 \right .\nonumber\\
&+&\left .  {2\over 3}\left (\eta_{Z'}^{-3/21}\eta_{b}^{-3/23}\eta_D^{-3/25} -
\eta_{Z'}^{-30/21}\eta_{b}^{-30/23}\eta_D^{-30/25}\right ) {1\over 4}
a_{ij}b_{ij}(1- 6\epsilon)\right ).
\end{eqnarray}
In this result $\eta_{Z'} = \alpha_s(m_{Z'})/\alpha_s(m_t)$,
$\eta_b = \alpha_s(m_t)/\alpha_s(m_b)$, and $\eta_D =
\alpha_s(m_b)/\alpha_s(m_D)$ and the rest of the notation is
defined  in Ref.~\cite{He:2006bk}. We will take $M_{Z^\prime}\sim
500$~GeV for illustration.

There are different ways in which non-universal $Z^\prime$ couplings to quarks can arise. Here we will
use a model resulting from a
variation of Left-Right models \cite{He:2002ha} that is motivated by the apparent anomaly in the $A^b_{FB}$
measurement at LEP~\cite{Chanowitz:2001bv}. In this model, some relevant $Z'$
couplings to fermions are given by
\begin{eqnarray}
L &=& {g\over 2 \cos\theta_W} \tan\theta_W (\tan\theta_R +
\cot\theta_R) \cos\xi_Z Z'_\mu \nonumber\\
&\times&\left ( V^{d*}_{Rbi}V^d_{Rbj} \bar d_{R_i}\gamma^\mu
d_{Rj} - V^{u*}_{Rti}V^u_{Rtj} \bar u_{R_i}\gamma^\mu u_{Rj} +
\bar \tau_R \gamma_\mu \tau_R - \bar \nu_{R\tau} \gamma_\mu
\nu_{R\tau}\right ).
\label{fcnczp}
\end{eqnarray}
Here $\theta_W$ is the usual electroweak mixing angle, $\theta_R$
parameterizes the relative strength of the right-handed
interaction, $\xi_Z$ is the $Z-Z'$ mixing angle and
$V^{u,d}_{Rij}$ are the unitary matrices that rotate the
right-handed up-(down)-type quarks from the weak eigen-states to
the mass eigen-states. In order to explain the anomaly in
$A^b_{FB}$, an enhanced $Z'$ coupling to third generation fermions
is required, and this is paramterized by a large $\cot\theta_R$.
In Eq.~\ref{fcnczp} we have only written those terms that are
enhanced by $\cot\theta_R$ and refer the reader to
Ref.~\cite{He:2002ha} for more details. Additional
constraints that are placed on the model by measurements at LEP-II can be approximately summarized by the relation:
$\tan\theta_W\cot\theta_R\frac{M_W}{M_{Z^\prime}}\sim 1$
\cite{He:2002ha}.

In this model, the tree-level exchange of a $Z^\prime$ can generate mixing in both down-type and up-type quark sectors.
We have previously shown that the model can generate a large $B_s$ mixing and this suggests that  large $D$
mixing may also occur.
To see why a large $D-\bar D$ mixing is possible in general, let
us compare the tree-level contribution of $Z'$ exchange to $B-\bar
B$ and $D-\bar D$ mixing. In this case $a_{ij} =0$ and $b_{ij}$
is given respectively by
\begin{eqnarray}
&&\mbox{$B_q-\bar B_q$ mixing}:\;\; b_{sq} = \cos\theta_W \tan
\theta_W (\tan\theta_R + \cot\theta_R) \cos\xi_Z
V^{d*}_{Rbb}V^d_{Rbq},\nonumber\\
&&\mbox{$D^0-\bar D^0$ mixing}:\;\;b_{cu} = \cos\theta_W \tan
\theta_W (\tan\theta_R + \cot\theta_R) \cos\xi_Z
V^{u*}_{Rtc}V^u_{Rtu}.
\end{eqnarray}
It is clear that if $V^{u*}_{Rtc}V^u_{Rtu}$ is of the same order as
$V^{d*}_{Rbb}V^d_{Rbq}$, large $D^0-\bar D^0$ mixing
will result.

As mentioned before,  the SM contribution to $B_s-\bar B_s$ mixing is
slightly below the central experimental value allowing a $Z'$ contribution.
Under this scenario, it was shown
in Ref.~\cite{He:2006bk} that the parameter
$|V^{d*}_{Rbb}V^d_{Rbs}|$ is of order $10^{-4}$. For example, if one includes the
large one-loop contribution that is present for $B_s$ mixing, $a_{bs}$, and fixes its value to
reproduce $K^+ \to \pi^+ \nu \bar\nu$ \cite{Anisimovsky:2004hr}, then
\begin{equation}
0.0005  \sim \left|V^{d*}_{Rbb}V^d_{Rbs}\right| \sim 0.0009.
\label{bsres}
\end{equation}

For the case of $D^0-\bar D^0$ mixing, the one-loop induced $a_{cu}$ is very small
(suppressed by $m^2_b/m^2_W$) compared to $b_{cu}$ and can be neglected. This is
unlike what happens for $B$ and $K$ mixing, where $a_{bs}$ and $a_{ds}$ must be
considered \cite{He:2006bk}.

For our numerical analysis, let us first consider the CP conserving case with $\phi=0$.
Taking $\Gamma_{D^0}$ from the particle data
group \cite{pdg}, $\epsilon = 1$, $f_D\sqrt{B_D} = 200$ MeV, and
using the approximate constraint $m_W/m_{Z'} (\tan\theta_W\cot\theta_R) =1$, we obtain
\begin{eqnarray}
x = {2M_{12}\over \Gamma_{D^0}} = 1.9 \times
10^{5}\ |V_{Rtc}^{u*}V^u_{Rtu}|^2.
\end{eqnarray}
Requiring that this $x$ due to the $Z^\prime$ exchange be at most equal to the measured $x$ (the average from Ref.~\cite{Ciuchini:2007cw}) , we
find that $| V^{u*}_{Rtc}V^u_{Rtu} |$ is bound by
\begin{eqnarray}
| V^{u*}_{Rtc}V^u_{Rtu} | \lsim 2.0\times 10^{-4}.
\label{rangeres}
\end{eqnarray}

This is the tightest bound on FCNC in the
up-quark sector for these models that we can obtain at present.
Of course, if the standard model (long distance) contributions to $x$ could be precisely quantified, then the constraints on new physics would be much tighter. To gauge the significance of this constraint we note that if
the matrix $V^{u}_{Rij}$ has a similar hierarchy to the one exhibited by the
CKM matrix, the value for $|V_{Rtc}^{u*}V_{Rtu}^u|$ would be of
order $10^{-4}$. Notice also that the upper bound for
$V_{Rtc}^{u*}V_{Rtu}^u$ is similar to the range allowed for
$|V^{d*}_{Rbb}V^d_{Rbs}|$ in $B_s - \bar B_s$ mixing.

With a non-zero CP violating phase $\phi$, the upper bound will be different in general. In the $Z^\prime$ model  the phase
$\phi$, appears as a complex $V^{u*}_{Rtc}V^u_{Rtu} =
|V^{u*}_{Rtc}V^u_{Rtu}|e^{i\phi/2}$. A numerical study allowing $x$, $y$ and $\sin\phi$ to vary independently in their one-sigma ranges results in essentially the same bound, Eq.~\ref{rangeres}, for the whole range of $\sin\phi$.

We now assume that $D-\bar D$ mixing is indeed due to FCNC induced by $Z^\prime$
exchange and explore the consequences for rare $D$ and top decays
which are also governed by $V^u_{Rij}$.
The $Z^\prime$ couplings are
enhanced by $\cot\theta_R$ only for third generation fermions, so
the most promising rare decay would be $D^0 \to X_u \nu_\tau
\bar\nu_\tau$. Since the standard model rate for this process is
too small to be observed \cite{Burdman:2001tf}, we consider only
the new physics contribution from tree-level $Z^\prime$ exchange.
We further work at the spectator quark level to obtain
\begin{eqnarray}
{\cal B}(D^0 \to X_u \nu \bar\nu) &\approx & \frac{G_F^2
m_c^5}{768 \pi^3 \Gamma_{D^0}} |V_{Rtc}^{u*}V_{Rtu}^u|^2
\left(\tan\theta_W\cot\theta_R\frac{M_W}{M_{Z^\prime}}
\right)^4\nonumber\\ &\lsim&  3 \times 10^{-10},
\end{eqnarray}
still too small to be observed in the near future.

Other rare decays that involve only second generation fermions are even smaller.
For example, for the $Z^\prime$ contribution to $D^0\to \mu^+\mu^-$ we find
\begin{eqnarray}
{\cal B}(D^0 \to \mu^+\mu^-) &\approx & \frac{G_F^2 m_D m_\mu^2F_D^2}{16 \pi \Gamma_{D^0}}
|V_{Rtc}^{u*}V_{Rtu}^u|^2 \left(\tan\theta_W\cot\theta_R\frac{M_W}{M_{Z^\prime}}
\right)^4 \tan^4\theta_R \nonumber \\
&\lsim & 4 \times 10^{-15} \left(\frac{\tan\theta_R}{0.1}\right).
\end{eqnarray}
Similarly, charged $D$ rare decays also remain too small for observation.

In contrast with the rare $D$ decay, non-universal $Z'$ exchange can significantly
enhance rare top-quark decays because more third generation fermions are involved. For example we find,
\begin{eqnarray}
{\cal B}(t \to c \tau \bar \tau) &\approx
& \frac{G_F^2 m_t^5}{768 \pi^3 \Gamma_{t}}
|V_{Rtt}^{u*}V_{Rtc}^u|^2
\left(\tan\theta_W\cot\theta_R\frac{M_W}{M_{Z^\prime}}
\right)^4\nonumber\\
&=& 4.3\times 10^{-4} |V_{Rtt}^{u*}V_{Rtc}^u|^2, \nonumber \\
{\cal B}(t \to c b\bar b) &\approx & 3\frac{G_F^2 m_t^5}{768 \pi^3
\Gamma_{t}} |V_{Rtt}^{u*}V_{Rtc}^u|^2|V_{Rbb}^{d}|^4
\left(\tan\theta_W\cot\theta_R\frac{M_W}{M_{Z^\prime}}
\right)^4\nonumber\\
& =& 1.3\times 10^{-3}|V_{Rtt}^{u*}V_{Rtc}^u|^2|V_{Rbb}^{d}|^4,
\end{eqnarray}
and ${\cal B}(t \to c \tau \bar \tau)= {\cal B}(t \to c \nu_\tau \bar \nu_\tau)$.

Rare $t\to c (\nu \bar \nu, \tau\bar\tau, \bar b b)$ decays may
offer some hope for detection depending on the value of
$V_{Rtt}^{u*}$ and $V_{Rtc}^u$. The largest branching ratios can
be  of order $10^{-3}$, which may be observable at LHC and ILC
\cite{lhcilc}. If instead of taking the maximum possible mixing
angles one scales the matrix elements according to the CKM
hierarchy, the natural size for $V^u_{Rtt}$ is of order 1 and for
$V^u_{Rtc}$ is of order $4\times 10^{-2}$. This would lead to
${\cal B}(t \to c b\bar b) \sim 10^{-6}$ which is probably too
small to be observed. It is also possible to search for the new $Z^\prime$ through flavor conserving processes at the LHC and this has been considered previously in the literature \cite{Han:2004zh}.

Up to now we have considered  a class of non-universal $Z'$
models that is motivated by the measured  $A^b_{FB}$. In
this class of models, the quantity
$\tan\theta_W\cot\theta_R\frac{M_W}{M_{Z^\prime}}$ is constrained to
be near one and this allows us to directly constrain
the matrix element $V^{u*}_{Rtc}V^u_{Rtu}$. In a general
$Z'$ model, one cannot do this, and instead  has to work with
the parameter $b_{cu}$ and the $Z^\prime$ mass. The constraint we have obtained for $|V^{u*}_{Rtc}V^u_{Rtu}|$
applies in general to
$(|b_{cu}|m_Z/m_{Z^\prime})$.

To summarize, we have investigated the implications of recent
results on $D^0-\bar D^0$ mixing for a class of non-universal
$Z^\prime$ models. We pointed out that the contributions of these
models to $B_s-\bar B_s$ mixing allowed by the measurement of
$\Delta M_{B_s}$, suggest that the models could also induce
$D-\bar D$ mixing at current levels of sensitivity. Within the allowed
parameter space, the CP violating semileptonic asymmetry $|a|$ can be as large as
$\sim 30\%$. We find that $D$ mixing constrains the models in such a way
that rare $D$ decays are unobservably small. Rare top decays are
larger but most likely too small to be observed. Our numerical analysis was based on requiring that the new physics contribution to $x$ be at most as large as the measurement (at $1-\sigma$). Clearly this can be improved if the SM contribution to $x$ is understood better.

\noindent {\bf Acknowledgments}$\,$ The work of X.G.H. was
supported in part by the NSC and NCTS. The work of G.V. was
supported in part by DOE under contract number DE-FG02-01ER41155.

\end{document}